\documentclass[useAMS,usenatbib]{mn2e}

\usepackage[dvips]{graphicx}
\usepackage{graphics}
\usepackage{psfig}
\usepackage{amssymb}
\input{epsf}

\title[On the variability of the iron K$\alpha$ line in Mrk 841]{On the  variability of the  iron line in Mrk 841}
\author[A. L. Longinotti, K. Nandra, P. O. Petrucci, P. M. O'Neill]{A. L. Longinotti$^{1}$\thanks{E-mail:
all@imperial.ac.uk}, K. Nandra$^{1}$, P. O. Petrucci$^{2}$, P. M. O'Neill$^{1}$
\\
$^{1}$Astrophysics Group, Imperial College London, Blackett Laboratory, Prince Consort Road, London SW7 2AZ\\
$^{2}$Laboratoire d'Astrophysique de Grenoble, BP 43, 38041 Grenoble
  Cedex 9, France}
\begin{document}

\date{Accepted}

\pagerange{\pageref{firstpage}--\pageref{lastpage}} \pubyear{2004}

\maketitle

\label{firstpage}

\begin{abstract}
Petrucci et al. (2002; hereafter P02) have reported extraordinary behaviour of the iron K$\alpha$  line in the type 1 AGN Mrk 841. At the {\it XMM-Newton}/EPIC resolution, a narrow line was observed in a short observation, which then apparently disappeared around 0.5 days later when another observation was performed. The limits are such that the line cannot both be narrow, and variable at this level, as the light-travel time places the material at a radius where broadening by the black hole's gravitational field is inevitable. Adding an additional observation taken before the P02 data, we present a different interpretation of the apparent variability.  The data support the hypothesis that the line in fact varies in width, not only in  flux.  Rather than showing no line, and despite the strong Compton reflection component, the final observation exhibits broad emission at $\sim 6.2-6.4$~keV, with a width $\sigma \sim 0.8$~keV, typical of a relativistic accretion disc.  We suggest that the apparently narrow line in the early observations arises from local illumination by a flare inducing an hotspot  in the inner disc, which then becomes progressively broadened as the disc rotates.  
\end{abstract}


\section{Introduction}

The fluorescence iron K$\alpha$ line is a well-established 
feature in the hard X-ray spectra of Active Galactic Nuclei (AGN) (Nandra \& Pounds 1994).
The properties of this line can provide valuable information on
the regions surrounding the AGN. For example, narrow Fe K$\alpha$ emission
can be associated with optically thick material 
at large distance from the nuclear region (Ghisellini, Haardt \& Matt 1994;
Krolik, Madau \& Zycki 1994). Such narrow lines are commonly
observed in high resolution spectra taken with the {\it Chandra}
HETG (Yaqoob et al. 2001; Yaqoob \& Padmanabhan 2004). 
Broad lines were often observed in the {\it ASCA} spectra of Seyfert
galaxies  (Tanaka et al 1995;
Nandra et al. 1997), and they were interpreted 
as the result of Doppler  broadening and gravitational redshift 
in the proximity of a supermassive black hole
(Fabian et al. 1989). The existence of strong narrow components revealed
by {\it Chandra} has caused the modelling of the broad emission seen by {\it ASCA}
to become ambiguous in some cases, where complex absorption can be invoked
to explain some - and in some objects all - of the broad residuals (e.g. Pounds \& Reeves 2002, Reeves et al. 2004). On the other hand, the best example of a broad
accretion disc line, MCG$-$6$-$30$-$15, seems to survive all attempts at modelling the
spectrum with a complex continuum (Fabian et al. 1995, 2002; Vaughan \& Fabian 2004).
Nevertheless, given the inherent ambiguity in the spectral modelling, it is clearly important to seek other clues and indications of the origin of the emission near the iron K-complex. Line variability is the most obvious complement to the spectra. 

Mrk 841 is a type 1 AGN at {\itshape z} = 0.0365  (Veron-Cetty, 2001).
Several previous observations performed with different X-ray observatories ({\it EXOSAT, Ginga} and {\it  ROSAT})  have reported on its spectral variability (George et al, 1993, Nandra et al 1995).
Evidence for a broad iron emission line has been reported in one of two {\it ASCA} spectra reported by
Nandra et al. (1997), and Weaver, Gelbord \& Yaqoob (2001) have reported marginal
evidence for variability of the line. The presence of a Compton reflection
component  has been studied and 
confirmed by {\it BeppoSAX} (Bianchi et al, 2001, 2004) and {\it XMM-Newton} observations
by P02. The most interesting result from the P02 observation, however, 
was the detection of apparent rapid ($\sim$0.5 days) variability   
of an Fe K$\alpha$ line which was also rather narrow. P02 showed that the small
line width observed by {\it XMM-Newton} - if due to Keplerian motions implying a distant origin for the line - was inconsistent with the rapid variability for any reasonable black hole mass
(Woo \& Urry 2002). This emphasises the ambiguity inherent in static spectral modeling as described above. The variability constraint in Mrk 841 is stronger than the spectral constraint, as the former relies only on causality and the Eddington limit, while the latter depends on the uncertain velocity field of the line emitting material. 

In this Letter an additional {\it XMM-Newton} exposure has been added to 
the one published in P02, giving better statistics
and longer temporal coverage. This permits an alternative
intepretation of the P02 data, which may reconcile the rapid line
variability with its small width. 
 
\section{Observations and data reduction}
Mrk 841 was observed by {\it XMM-Newton} between the 13th and 14th of January 2001, in 3 different 
pointings, referred to as Obs I, II, III in chronological order (observations II and III are
those published in P02).
The respective durations are 9.3, 11.9 and 14.4 ks and I and II are contiguous, whereas 
there is a $\sim$ 43 ks gap between II and III.
The data reported here come from the EPIC instrument:
both of the pn and MOS cameras were in small window mode, with only the MOS2 data available   
for analysis as the MOS1 was in FAST UNCOMP mode.
Observations II and III were performed with the thin filter while a medium filter was used in 
Obs. I. 

The data were reduced using the {\it XMM-Newton} Science Analysis Software, version 5.4.1.
No background flares were found in the light curve, and having taken into account the dead
time, the filtered event files contained   5.9,7.6 and 9.2 ks of  good exposure   
in the pn detector, for Obs I, II and III, respectively.
Only the pn analysis is described in the following.
The data from MOS2 have been analysed: they confirm the following results but they are not included 
as the constraints are less strong than those given by the pn. 
Source events have been collected in all 3 data sets from a circular region with a 40$^{\prime\prime}$  radius centred on the source, while background events have been collected from
a larger region of 56$^{\prime\prime}$  radius, chosen in a nearby source-free portion 
of the detector.
Single and double events are selected in the spectra yielding a 
number of source counts between 2--10 keV
of 9374 counts (Obs I), 13495 (Obs. II) and 17510 (Obs III). 
Pulse-height spectra were binned in order to have at least 20 counts/bin which allows
the use of  $\chi^2$ minimization for spectral fitting.
\begin{figure}
\begin{center}
\psfig{figure=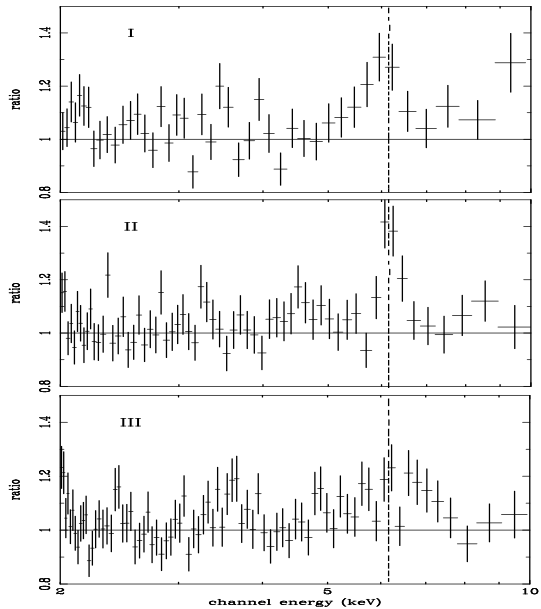,height=8cm,width=.4\textwidth}
\caption{Data to model ratios of the three data sets (observer's frame), 
fitted with a power law with $\Gamma$= $\sim$ 1.9, excluding the 5-7 keV energy 
channels: the dashed line marks the rest-frame energy of the neutral Fe K$\alpha$ line 
6.4 keV. The line profile is narrow in the top and middle panels, while it is 
considerably broadened in the bottom one.}
\end{center} 
\end{figure}

\section{Spectral analysis}
 Throughout the analysis errors are quoted  at  $\bigtriangleup \chi^2$= 4.61 (90 per cent of confidence for two interesting parameters). We began by fitting each spectrum individually in the 2-10  keV rest frame range with a model consisting of a power law and the Galactic column density (N$_H$=2.33 $\times$ 10$^{20}$~cm$^{-2}$). 
The power law slopes derived
for each observation were consistent with each other:
1.94$^{+0.06}_{-0.08}$ (I), 1.95$^{+0.04}_{-0.04}$ (II) and 
1.98$^{+0.04}_{-0.06}$ (III).\\
Fig. 1 shows the 2--10 keV spectra plotted as the ratio of the 
observed pn data to a
simple power law model, fitted to the spectra ignoring the 5-7 keV channels.
Residuals at $\sim$ 6.4 keV rest-frame (dashed line) reveal the 
presence of an Fe K$\alpha$ line.

Fitting a narrow  line to these individual spectra
confirms the result of P02, i.e. that the apparent flux drops in Obs. III
If the line width is left free to vary, however, a
different scenario is implied. As is evident from Fig. 1, the line
in Obs. III does in fact appear to be present, but is much broader than in the
other two datasets. To better constrain this phenomenon, and
since the photon indeces derived by fitting the spectra individually
are consistent with each other, we assume the broadband spectral 
shape to be constant.
Therefore, in the following we perform the spectral analysis for all 
the three observations {\it  simultaneously},
allowing the cross-normalizations (and often the line parameters, as 
described) to be free. 

Adding a gaussian line with the width free to vary improves the fit with  
$\bigtriangleup \chi^2$= 74, with respect to a simple power law model.
The rest-frame line energies are  6.29$^{+0.25}_{-0.41}$ keV,  6.37$^{+0.11}_{-0.07}$ keV and 6.22$^{+0.66}_{-0.74}$ keV  for I, II and III respectively, but the line widths show some difference, with best fitting values  0.25, 0.13 and 0.88 keV for the three observations. 

The parameters  in Obs. I and II are consistent with each other  and
these two pointings are nearly  consecutive in time, followed by a
long gap (see previous section).
For this reason therefore, the line parameters have been tied
  together  combining (Obs. I + II), in  order to compare with
the broad line features seen in Obs. III.\\
The best fitting parameters are summarised in Table 1.
In Fig. 2 we plot the confidence contours of the line widths versus 
the line fluxes:
it is clear from this picture that the primary variations is not in
flux (which is consistent with a constant within the error bars)
but the line profile - parameterized here by a broadening of the
gaussian width.

It is this interesting and unusual possibility that we investigate
further below. We have therefore continued to
tie the parameters of (I+II) together in the fits below, and where
appropriate have also tied parameters between Obs. (I+II) and
Obs. III.
In the following models the line parameters are kept tied so to be the same 
in (I+II);  also  the energies have been tied to have the same value in all the observations.

\begin{figure}
\begin{center}
\psfig{figure=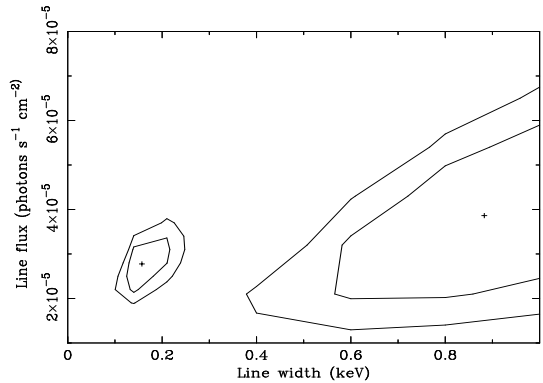,height=5cm,width=.4\textwidth,angle=-90}
\caption{Confidence contours at 68.3 and 90 per cent level of confidence for the line widths and fluxes when data are fitted with table 1 model. The left contour refers to (I+II) and the right 
contour refers to III alone.}
\end{center} 
\end{figure}

The large line broadening in observation III ($\sigma \sim$ 0.88 keV) can be the effect of reflection in the inner accretion disc which, if optically thick, will also introduce a Compton reflection component into the spectrum.  We have modelled this using the reflection model of Magdziarz \& Zdziarski (1995; called "PEXRAV" in the XSPEC spectral fitting package),
with the disc inclination parameter {\it i} set to be 30$^{\circ}$. 
We fit this together with a neutral Fe K line,
and  account  for the relativistic effects in the vicinity 
of the central black hole using the blurring code described by Fabian et al. (2002), in 
Kerr metric (Laor 1991), which we refer to as KDBLUR. 
In this way the reflection  continuum gets smoothed and smeared 
and the line broadened  according to an emissivity law as a function of radius of the form
{\itshape j} $\propto$ r$^{-q}$, where {\itshape j} is the line emissivity and r is the radius.
The inner and outer radii have been fixed to $r_{\rm in}=1.24 r_{\rm g}$ and
$r_{\rm out}=100 r_{\rm g}$, respectively. 
The model is shown in the left panel of Fig. 3 and best fitting parameters 
are quoted in Table 2. 
The derived emissivity index in (I+II) is q=$-$10, the lowest value permitted by the model, 
indicating that the line emission  in (I+II) is concentrated 
at large radii. On the other hand the line emissivity in III has q=2.1, which is much more typical 
(Nandra et al. 1997), and indicates that the line is concentrated in the inner regions.   
The upper limit on the inclination angle obtained in KDBLUR  implies  the disc is consistent
with being viewed  at a low inclination ($<$ 25$^{\circ}$). 
The confidence contours in the right panel of Fig. 3  show that the reflection component 
is rather strong, since the  value of the solid angle of the reflecting material seen by the X-ray source, R=$\Omega$/2$\pi$ $\sim 2$, 
is high, R = 2.1$^{+1.26}_{-0.82}$. 
This is in good agreement 
with the data in P02, despite the fact that they did not incorporate the relativistic blurring, and
that their constraints were improved by the inclusion of {\it BeppoSAX} high energy data
(see also in Bianchi et al 2001).   
Despite the large reflection, the line equivalent widths are only $\sim 120$ eV, possibly implying 
a small Fe underabundance for the observed values of $\Gamma$ and R  in the 
reflection model of George \& Fabian (1991).

\begin{table*}
\begin{minipage}{14cm}
\scriptsize
\begin{center}
\caption{Best fitting parameters for (power law + zgauss) model 
with the first two observations tied together. The energies are quoted in the source 
rest-frame}
POWER LAW+ZGAUSS\\
\begin{tabular}{c c c c c c c }
 \\ \hline\hline
\\
Obs. & $\Gamma$ & E & $\sigma$  & EW &  Flux (line)   &                       $\chi^2$/dof\\
(Chr. order) &   & (keV) &  (keV) & (eV) & (10$^{-5}$ ph cm$^{-2}$ s$^{-1}$)         &    \\       
\hline
\\
I + II &  1.92$^{+0.04}_{-0.03}$ &  6.36$^{+0.08}_{-0.06}$ & 0.15$^{+0.10}_{-0.08}$ & 174$^{+64}_{-61}$ & 2.77$^{+0.83}_{-0.72}$  & 1369/1423\\
\\  
III  &idem   & idem  & 0.88$^{+0.98}_{-0.54}$ & 237$^{+246}_{-162}$ & 3.86$^{+2.74}_{-2.11}$ & \\ 
\hline\hline
\end{tabular}
\end{center}
\end{minipage}
\end{table*}

\begin{figure*}
\begin{center}
\centerline{
\resizebox{7.2cm}{!}{\psfig{figure=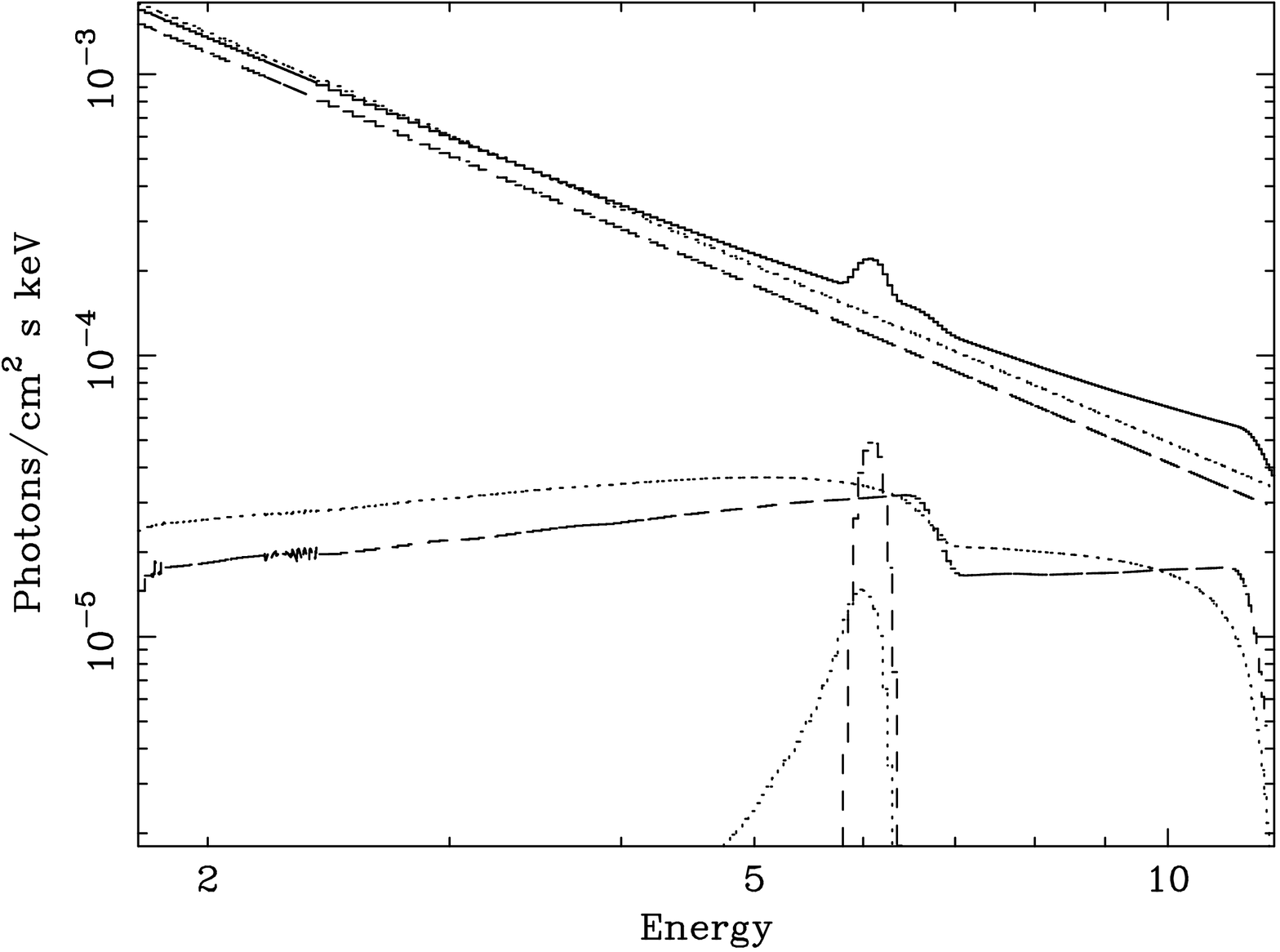,angle=-90}}
\hspace{0.3 cm}
\resizebox{7cm}{!}{\psfig{figure=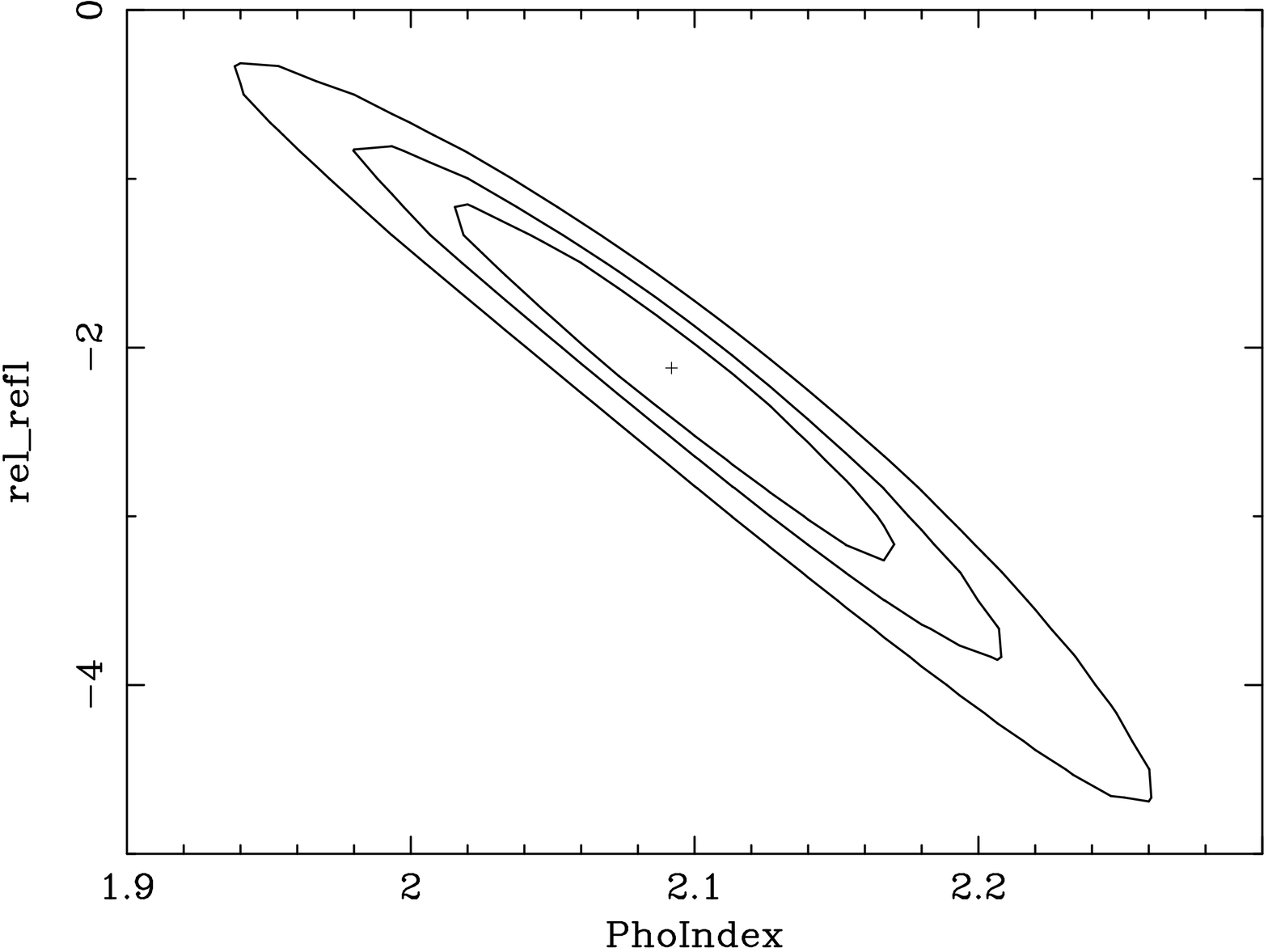,angle=-90}}
}
\caption{Left panel: table 2 best fitting reflection model.
The blurring code acts only on the line and the reflection continuum, yielding 
a very broadened line in Obs III.  Right: 68.3, 90 and 99 per cent confidence contours of the photon index versus the reflection fraction parameter R in the same fit; the value of $\sim$ 2.1 indicates a strong component, still consistent with the previous ones observed by {\it XMM-Newton} and {\it BeppoSAX}}.   
\end{center} 
\end{figure*} 

\begin{table*}
\begin{minipage}{14cm}
\scriptsize
\begin{center}
\caption{Best fitting parameters for the reflection model adopted; the radii 
in the KDBLUR blurring routine are fixed to r$_{in}$=1.24 r$_g$ and r$_{out}$=100r$_g$. The line emissivity scales with radius according to {\itshape j }$\propto$ r$^{-q}$.}
POWER LAW+ (PEXRAV + GAU)*KDBLUR\\  
\begin{tabular}{c c c c c c c}
 \\ \hline\hline
\\
Obs. & $\Gamma$ & R  & {\itshape i} & q &  EW &$\chi^2$/dof     \\
(Chr. order) &   &   &   degrees & & (eV) & \\
\hline
\\
I + II &  2.09$^{+0.14}_{-0.13}$ &  2.10$^{+1.26}_{-0.82}$ & $<$ 25   & -10 & 113$^{+60}_{-54}$ & 1368/1423 \\
\\  
III  & idem   & idem  & idem & 2.08 & 70$^{+123}_{-70}$ & -\\ 
\hline\hline
\end{tabular}
\end{center}
\end{minipage}
\end{table*}
 
A gross change in a power-law line emissivity law seems an artificial and unlikely explanation  for the change in line profile. Arguably a more plausible alternative is that the line arises from illumination by discrete flares acting over a limited radial range.  A  large variation in the line width could be explained if the flaring regions were
concentrated at large radii in (I+II), but occurred much closer to the black hole in III.
To test this, we set the outer radius in KDBLUR to be a factor of 1.1 larger than the inner radius, defining a ring on the disc where the line originates; here the emissivity index has been fixed to q= 3 and the inner radius has been left free to vary.
With these conditions, there is a small improvement of $\bigtriangleup \chi^2$=5 in the fit 
(see Table 3).
The reflection fraction is slightly diminished, R $\sim$ 1.6, better reconciling with a solar Fe 
abundance for the observed EW (between $\sim$ 110-130 eV).
With such a  fit the emission regions must be disjointed.
 The 90 per cent lower limits on the inner radii for (I+II) is 100 r$_g$ - i.e. the emission comes from outside
this radius.
On the other hand, the broad line in Obs. III is constrained in a very inner ring, at less than 20 r$_g$.    
As it is  discussed in detail later, however, it is also possible that the 
line emissivity is not well modeled by a power law, or even circularly symmetric, so the 
results of these fits could be quite misleading.
\begin{table*}
\begin{minipage}{14cm}
\scriptsize
\begin{center}
\caption{Best fitting parameters of the reflection model with  r$_{out}$= 1.1 $\times$r$_{in}$ and
the emissivity index q=3. The $\chi^2$ is by far the best obtained. The energies are in the source rest-frame.}
PO + (PEXRAV + GAU) *KDBLUR  
\begin{tabular}{c c c c c c c c}
 \\ \hline\hline
\\
Obs. & $\Gamma$ & R & {\itshape i} &  E$_{line}$ &  EW & r$_{in}$ & $\chi^2$/dof \\
(Chr. order) &   &  & degrees  &  (keV) & (eV) & r$_g$ &  \\       
\hline
\\
I + II &  2.05$^{+0.13}_{-0.07}$ &  1.62$^{+0.39}_{-0.35}$ &  $<$ 26 & 6.41$^{+0.06}_{-0.10}$ & 114$^{+50}_{-48}$ &  $>$ 100 & 1363/1420 \\
\\  
III  &idem   & idem  & idem & idem &   135$^{+60}_{-56}$ & $<$ 20 & -  \\ 
\hline\hline
\end{tabular}
\end{center}
\end{minipage}
\end{table*}

\section{Discussion}
Three {\it XMM-Newton} spectra of Mrk 841   have been analysed. The main results are:
\begin{itemize}
\item The 2--10 keV continuum is well fitted by a power law of $\Gamma$=2.05$^{+0.13}_{-0.07}$. 
\item The presence of a fluorescent Fe K$\alpha$ line 
consistent with neutral reflection is confirmed in all the observations.
\item Previous analysis of roughly the same data claimed evidence 
for  variability of a factor of $\sim$ 1.6 in the narrow line flux and 
of $\sim$ 10--20 per cent in the continuum (P02).
The present results indicate instead a much lower variation in the line flux, referring 
to the best fitting values, but they are still consistent with   variability if the error bars are taken into account (see table  1). Anyway, a clear {\itshape broadening} of the line in observation III  is detected with 99 per cent  confidence.
\item The strong reflection component reported both by P02 and  Bianchi et al. (2001),
is confirmed, with a value of the solid angle of the reflecting material seen by the X-ray source 
R=$\Omega$/2$\pi$ $\sim 2$.
Accretion disc modelling of the broad line with  low inclination, $<$ 26$^{\circ}$, and
 line equivalent of $\sim$ 120 eV, would suggest somewhat less than solar abundance  given the strong reflection component.
\item If modelled by a ring of emission in the accretion disc, the line in the two different time intervals is emitted in  two very different 
regions of the putative accretion disc, one placed within $\sim$ 20 r$_g$ from the spinning black hole and one at  $>100$ r$_g$.
\end{itemize}     

If our interpretation of the variability - that the  line changes in profile, rather than flux - is correct, it must also be reconciled with a reasonable physical interpretation and appropriate time scales in the disc. The key observational facts are that, 
during the first 21 ks, the line width is 0.15$^{+0.10}_{-0.08}$ keV (I+II), then the detectors  
are switched off for $\sim$ 43 ks before being operated again for the last $\sim$ 15 ks (III).
In this last exposure the line is broadened by $\bigtriangleup$$\sigma$$\sim$ 0.73 keV. Is this consistent with any reasonable time scale in the system? 

Time-scales  quoted in the following are in units of  M$_8$= 10$^8$M$_{\odot}$ consistent with the black hole mass  estimated by Woo \& Urry (2002).  The black hole mass was estimated by these authors  using the Broad Line Region size derived from the monochromatic luminosity
at 5100$\AA$ in ergs s$^{-1}$, R$_{BLR}\propto L_{5100}^{0.7}$, which  combined with  the BLR clouds velocity estimated by the mean Full Width Half Maximum of the optical spectral lines in km s$^{-1}$, leads to 
M$_{BH}$= 4.817 $\times$ [$\frac{\lambda L_{\lambda}(5100\AA)}{10^{44} ergs  s^{-1}}$]$^{0.7}$ FWHM$^2$~M$_{\odot}$. We note, however, that this mass estimate is subject to considerable uncertainty. 
\begin{figure}
\begin{center}
\psfig{figure=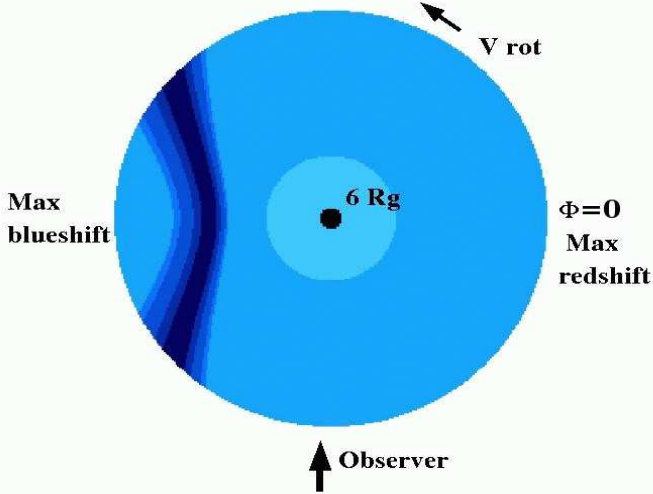,height=7.5cm,width=.55\textwidth}
\caption{A schematic map of the inner accretion disc showing the allowed region were a narrow line
 with parameters consistent with spectrum (I+II) could be emitted.
The different shades represents regions of increasing confidence. The darkest shade corresponds to the centroid energy 6.36$^{+0.03}_{-0.04}$ keV,
with errors within $\bigtriangleup\chi^2$=1, (68 per cent of confidence for one interesting parameter).
The lighter colours correspond respectively to errors of 90, 
 99, 99.73 per cent. We assume the Schwarzschild case so no line can be generated in the innermost
 region within 6 $R_{\rm g}$ (lightest shade).}   
\end{center} 
\end{figure}

The simplest interpretation consistent with our modelling of the line is that at the time of observations (I+II), the disc was mainly illuminated at large radii, and during observation III, the illumination was mainly in the central regions, and thus broadened. A similar interpretation for the variable line profile in MCG$-$6$-$30$-$15 has been given by Iwasawa et al. (1996).

 There are, however, important differences here. Firstly, the difference in width is much greater, with the narrow line in (I+II) implying dominant emission outside 100 r$_g$.  It seems difficult to imagine that X-ray flares dominating the emission at such a large radius would then dissipate completely 0.5 days later. 
In addition, the near-constancy of both the line
and continuum fluxes argues strongly against this interpretation (c.f. MCG$-$6$-$30$-$15): there is no good reason why flares at such different radii would have the same luminosity. 
One could imagine a flare beginning at large radii and propagating inwards, retaining roughly the same power. However, only at the light crossing time can a flare reflected 
at $\sim$ 100 r$_g$ travel to $\sim$ 10 r$_g$ on a time-scale consistent 
with our observations. The appropriate timescale is 
 $\bigtriangleup$$\tau$ = 4.4 $\times$ 10$^4$ M$_8$ s
having assumed that the matter travels at the speed of light. 
This possibility is also unlikely: the matter in the disc certainly 
is accreted  at a velocity much lower than {\itshape c} which 
makes the  $\bigtriangleup$$\tau$ longer than estimated above. 

Perhaps the most plausible interpretation of the data consists in a localized hotspot 
co-rotating with the disc.
Assuming keplerian rotation, t$_{kepl}$= 2.85 $\times$ 10$^5$ M$_8$ (r/10r$_s$)$^{3/2}$ s (Treves,
Maraschi \& Abramowicz 1988)
which leads to t$_{kepl}$$\sim$ 1 $\times$ 10$^5$ M$_8$ s for a complete orbit at 10 r$_g$,  t$_{kepl}$$\sim$ 0.46 $\times$ 10$^5$ M$_8$ s at 6 r$_g$ and 
$\sim$ 0.043 $\times$ 10$^5$ M$_8$ s at the last stable orbit r= 1.24 r$_g$.
Thus, the rotation time is consistent with the variability timescale as long as
the action occurs at a small radius. 

An isolated flare on the disc would illuminate a narrow range of azimuthal angle (and hence Doppler factor), resulting in a narrow line.
As demonstrated in Dovciak et al. (2004), the blue horn originated by an internal 
spot (6 r$_g$) in a disc observed at 30$^{\circ}$,  is very close to 6.4 keV and it can easily 
be misinterpreted as a narrow line originating further away.
These authors calculate and display line profiles integrated over 1/12 of the orbital period 
at the corresponding radius.
In a disc viewed at 30$^{\circ}$,  a line produced at r=10 and 20 r$_g$ 
can result in a narrow peak at almost 6.4 keV. 
The line in (I+II) is consistent with this scenario, being integrated over 
1/10 of the estimated orbital time at $\sim$25 $^{\circ}$  and $\leq$ 20 r$_g$.
      
 Over the Keplerian timescale, as the disc rotates, the line could be broadened 
 out as seen in observation III. 
The physical mechanism for the broadening mechanism and is
relationship to the rotation is unclear,
but it must be such that
the hotspot is well confined during the first part of the disk orbit and
subsequently, the emissivity is distributed over a larger region on the disc.
It is also possible that a shift of the spot inwards is contributing
to produce a broader line, even though it is unfortunately very 
difficult to place
tight constraints on the size or the position of the flaring region.
The model proposed  by Dovciak et al. predicts
that both line energy and flux should present some amount of variation,
over the orbital timescale; such variations cannot be tracked and
measured precisely
from the available spectra, but by means of a qualitative  comparison 
with fig. 2
in Dovciak et al., we can say that the $\bigtriangleup$E  and 
$\bigtriangleup$F inferred
from our observation, are consistent with what is theoretically expected.
This is perhaps unsurprising given the constraints: for example, when 
left free,
the line energy in Obs III spans a wide interval (6.25$^{+0.47}_{-0.71}$  keV).
It is therefore easily possible to find
  a region of parameter space consistent with the present observation 
as long as the spot
is located in the inner disc during the last part of the orbit, as 
defined by Dovciak in fig. 2 (i.e. Time = 0.7-1.0).

The phenomena discussed here are probably strongly related to the 
transient, narrow and shifted emission lines now reported for a number 
 of {\it XMM-Newton} and {\it Chandra} spectra
of AGN (Turner et al 2002, Guainazzi 2003,   Yaqoob et al. 2003, 
Turner et al.  2004, Porquet et al. 2004 ). 
The main difference in the case of Mrk 841 is 
that we see an apparent evolution of the line profile from narrow to broad. 
Although we cannot strongly rule
out the possibility with the present data that the broad component is 
actually present in the Obs. (I+II) spectrum, (the 90 per cent upper limit on 
the broad line in (I+II) is 162 eV), our 
interpretation of a profile variations seems the clearest
and most straightforward interpretation of the data.

There are though some effects to be taken in account which require
accurate tuning. 
When the line fluorescence takes place so close to the black hole, as it is case,          
the Fe K$\alpha$  photons are subject to a substantial gravitational redshift according to
E$^{\prime}$/E$_{\circ}$= (1-2r$_g$/r)$^{1/2}$ (Fabian et al. 1989).
At, e.g., 20 r$_g$  the line centroid would be shifted from the rest energy 6.4 keV to 
6.07 keV.
The Doppler shift must also be taken into account, and in particular if the flare
originates in the approaching part of the disc it will be blueshifted, counterbalancing
the gravitational redshift and, potentially, resulting in a narrow line close to 6.4 keV, as
observed in Obs. (I+II). 

To illustrate these effects, we have left the line energies untied 
in (I+II) and III and calculated
the appropriate gravitational and Doppler shifts in a Keplerian accretion disc for (I+II). 
Then,  we constructed a probability map of the expected line energy for each radius and azimuthal angle on the  disc. 
The map is displayed in Fig. 4: the maximum radius is 20 r$_g$ as we determine by fitting
spectrum III and we consider the
Schwarzschild case, so that no line originates within 6 r$_g$ (lightest shade).
The line energy is computed following the approximations 
in Fabian et al. (1989) and assuming the best fitting values in Table 3 
for the disc inclination and the upper limit on the emission radius.
The confidence regions refer to the errors on line energy quoted at
$\bigtriangleup\chi^2$= 1, 2.71, 6.63 and 9 for one interesting parameter,
corresponding to 68, 90, 99 and  99.73 per cent of confidence respectively.
The energy ranges are  illustrated in colour scale, starting from the 68 per cent region in the darkest shade (see caption for details).

The map demonstrates that the narrow line in Obs. (I+II) is consistent
with being emitted within the same radius found for the line in  Obs.  III, 
although from a restricted area on the inner disc. 
It is computed in  Schwarzschild metric excluding the region within 6r$_g$. However as the allowed region is outside this radius it should be fully consistent with the case of spinning black hole. 

A final caveat is to remember 
that the masses estimated from optical luminosity  suffer considerable 
uncertainty, as Woo \& Urry warned themeselves. 
The constraints on the emission region, and indeed on the line variability overall, would
be loosened considerably if this were taken into account. 
A prediction of our model is that the line seen in Mrk 841 during its "narrow" phase
is not truly narrow, and should have a finite and detectable width - albeit not
at the XMM/EPIC resolution.  New observations at high resolution with, e.g.
{\it Chandra}/HETG or {\it ASTROE-II} will be required to 
measure the width of the line  more precisely, and distinguish definitively
between the various options.

\section*{Acknowledgments}
This paper is based on observations obtained with the {\it XMM-Newton} satellite,
an ESA science mission with contributions funded by ESA Member States and USA.
We thank the {\it XMM-Newton} team for supporting the data analysis.
The authors wish to thank Prof. Andy Fabian for providing the relativistic 
blurring codes, Kazushi Iwasawa, Giovanni Miniutti and Stefano Bianchi for insightful 
discussion and the anonymous referee who provided fruitful suggestions. 
A.L.L. acknowledges financial support from the POE network at the Imperial College London and from the Angelo Della Riccia Foundation. 
P.M.O.acknowledges financial support from PPARC.

\end{document}